\documentclass[%
 aip,
 jcp,%
 amsmath,amssymb,
 floatfix,
 reprint,%
]{revtex4-1}

\usepackage[pdftex]{graphicx}%
\usepackage{dcolumn}%
\usepackage{bm}%
\usepackage{todonotes}

\begin{document}
\preprint{AIP/123-QED}

\title{
Accurate molecular dynamics and nuclear quantum effects 
at low cost by multiple steps in real and imaginary time: 
using density functional theory to accelerate 
wavefunction methods}%

\author{V. Kapil}
 \affiliation{Laboratory of Computational Science and Modelling, Institute of Materials, Ecole Polytechnique F\'ed\'erale de Lausanne, Lausanne, Switzerland}
\author{J. VandeVondele}%
 \email{joost.vandevondele@mat.ethz.ch }
\affiliation{Department of Materials, ETH Zurich, Wolfgang-Pauli-Strasse 27, CH-8093 Zurich, Switzerland}%

\author{M. Ceriotti}
 \email{michele.ceriotti@epfl.ch}
\affiliation{Laboratory of Computational Science and Modelling, Institute of Materials, Ecole Polytechnique F\'ed\'erale de Lausanne, Lausanne, Switzerland}%

\date{\today}%

\begin{abstract}
The development and implementation of increasingly 
accurate methods for electronic structure calculations
mean that, for many atomistic simulation problems, 
treating light nuclei as classical particles 
is now one of the most serious approximations. 
Even though recent developments have significantly
reduced the overhead for modeling the quantum nature
of the nuclei, the cost is still prohibitive 
when combined with advanced electronic structure methods. 
Here we present how multiple time step 
integrators can be combined
with ring-polymer contraction techniques (effectively, 
multiple time stepping in imaginary time) to reduce
virtually to zero the overhead of modelling 
nuclear quantum effects, while describing
inter-atomic forces at high levels of electronic structure theory. 
This is demonstrated for a combination of MP2 and 
semi-local DFT applied to the Zundel cation.
The approach can be seamlessly combined with other methods to
reduce the computational cost of path integral 
calculations, such as high-order factorizations of the
Boltzmann operator, or generalized Langevin equation thermostats.
\end{abstract}

\maketitle

Simulations of molecules and materials from first principles 
are constantly improving in accuracy and predictive power, thanks 
to the combination of the availability of more powerful 
high-performance computing (HPC) platforms and the development 
of more efficient techniques to treat the electronic structure
problem at high levels of theory. For example, the structure and dynamics of bulk liquid water can now be computed using correlated wavefunction methods \cite{delb+13jpcl,delb+15jcp2}, or quantum Monte Carlo
\cite{zen+15jcp}.
At the same time, 
these developments pose new challenges. To reach the
ultimate level of accuracy it is not sufficient to fully
account for exchange and correlations between electrons,
but it is also necessary to describe the quantum mechanical
nature of light nuclei, such as hydrogen. 
Modelling of nuclear quantum effects (NQEs) can be achieved
by path integral molecular dynamics (PIMD)
~\cite{feyn-hibb65book,chan-woly81jcp,parr-rahm84jcp},
that traditionally involves a very large overhead. 
The large computational cost can be traced
to the need of simulating multiple replicas of the physical
system and to the need for a reduced time step to properly integrate
the dynamics of the stiff, non-ergodic modes of the 
ring polymer Hamiltonian. Indeed, using brute force PIMD techniques, several hundred million core hours would be needed to faithfully model quantum effects in the above mentioned MP2 simulations of liquid water.

However, recent advances have contributed to dramatically reduce this
overhead, by using high-order expansions of the finite-temperature
density matrix\cite{jang-voth01jcp,pere-tuck11jcp}, by using a 
correlated-noise Generalized
Langevin Equation to induce frequency-dependent fluctuations mimicking
quantum effects~\cite{ceri+09prl2,ceri+11jcp,ceri-mano12prl}, or by selectively treating with different number of 
path integral replicas different parts of the system \cite{poma-dell10prl}
or different components of the potential\cite{mark-mano08jcp}. 
In a converged path integral simulation, adjacent
replicas in the ring polymer are held close together
by the presence of harmonic springs. Whenever it is 
possible to partition the inter-atomic potential into
a rapidly-varying and easily-computed part, and a 
harder-to-compute but slowly-varying part, one 
can also reduce the cost of evaluating the path energy
by a so-called ring-polymer contraction (RPC)
scheme~\cite{mark-mano08jcp,mark-mano08cpl}, that 
only evaluates the expensive 
component on a reduced number
of replicas. 

In fact, the idea of exploiting the presence of different
length scales in the interactions between atoms
has been recognized long since, and has been used to
accelerate the integration of classical equations of 
motion where the granularity of time is set 
by the stiffest part of the interaction~\cite{mart+92jcp}. 
This intuition can lead to significant speed-ups
when the rapidly varying part of the interaction
potential can be computed more efficiently than
the slowly-varying part -- which is e.g. the case
for long-range electrostatic interactions~\cite{zhou+01jcp},
or biasing potentials derived from complex order 
parameters~\cite{ferr+15jctc}. The partitioning 
of interactions in slow (smoothly-varying and expensive to compute) and 
fast (quickly-varying and inexpensive) components is less obvious when
first-principles calculations are considered, as individual terms are often hard to treat separately.
However, in this context, the vastly different costs between different levels of theory can be exploited.
This was first demonstrated in Ref.~\cite{guid+08jcp} where a combination of hybrid functionals and semi-local DFT was exploited in a multiple time-step (MTS) scheme. This early work was inspired by the combination of force-fields and DFT for presampling and biasing.\cite{vande-jcp00-free_calc, ifti-jcp00-class_pot, mcgrath-cpc05-isobar_isoth_mc}
Different variations on this theme, that 
all have the potential to be generally applicable, 
have been shown to be effective. For MTS schemes that
have been applied in an \emph{ab initio} context
include the combination of 
explicitly correlated methods on top of a baseline 
Hartree-Fock calculation~\cite{stee13jcp}, the 
application of a range-separated Coulomb operator or 
a divide-and-conquer strategy~\cite{lueh+14jcp},
the use of better-converged simulation parameters
as a correction, considered to be slowly 
varying~\cite{fate-stee15jctc}, or the combination of
MP2 and semi-local DFT\cite{delb+15jcp2}. The latter 
variant will be employed here, yielding a massive 
speedup by combining MTS and RPC.

\section{Combining multiple time step integrators and ring-polymer contraction}
MTS and RPC techniques can be combined fairly 
straightforwardly~\cite{geng15jcp}, and indeed, 
such a combination was used already in early classical
simulations based on ring-polymer contraction~\cite{habe+09jcp}.
However, to date no attempt has been reported to use 
RPC (alone or in combination with MTS) in the context 
of ab initio molecular dynamics, despite the fact that
this is a scenario in which obtaining an accurate estimate
of nuclear quantum effects at a reduced computational cost
would be particularly desirable. This is largely due to the 
considerable implementation overhead connected with combining 
sophisticated techniques for solving efficiently the
electronic-structure problem and the cumbersome
formalism that underlies PIMD and its RPC and MTS extensions.
Here we present a simple example that demonstrates a working
implementation of such combination, modeling nuclear quantum effects 
with a post-HF wavefunction description of the electronic structure. 
Our implementation relies on the modular assembly of i-PI \cite{ceri+14cpc} 
-- a Python interface in which we implemented MTS and RPC techniques -- and 
CP2K\cite{vand+05cpc} -- that sports a highly efficient periodic 
implementation of second-order M\o{}ller-Plesset (MP2) theory.
Commented example input files,
and a snapshot of our development code 
are provided in the 
supporting information.

We refer the reader to the relevant literature
for the theoretical foundations, and general considerations
regarding MTS and RPC techniques~\cite{mart+92jcp,mark-mano08jcp,mark-mano08cpl}. Here we will just 
briefly overview the main ideas, using as an example 
a path integral simulation of a flexible fixed-point-charge
model of water~\cite{habe+09jcp}. In the most straightforward
formulation, performing a PIMD simulation for a system 
composed of $N$ distinguishable particles with masses $m_i$ and positions 
$\mathbf{q}_i$ amounts to sampling at inverse temperature
$\beta_P=1/Pk_BT$
a canonical distribution consistent with the classical Hamiltonian
$H_P(\mathbf{p},\mathbf{q}) = H_P^0(\mathbf{p},\mathbf{q}) + V_P(\mathbf{p},\mathbf{q})$,
where the free ring-polymer Hamiltonian is
\begin{equation}
H_P^0(\mathbf{p},\mathbf{q}) = \sum_{i=1}^N \sum_{j=1}^P \left(\frac{[\mathbf{p}_i^{(j)}]^2}{2m_i} + \frac{1}{2} m_i \omega_P^2 [\mathbf{q}_i^{(j)} - \mathbf{q}_i^{(j+1)}]^2\right)
\label{eq:free-rp}
\end{equation}
and the physical potential term is
\begin{equation}
V_P(\mathbf{q}) = \sum_{j=1}^P V(\mathbf{q}_1^{(j)},...,\mathbf{q}_N^{(j)}).
\label{eq:tot-pot}
\end{equation}
The ring polymer is composed of $P$ replicas of the physical system
(positions $\mathbf{q}_i^{(j)}$) cyclically connected by springs with
frequency $\omega_P=P\beta\hbar$. To simplify the discussion of 
the combination of RPC and MTS, let us split the potential in 
a slow (S) and a fast (F) part 
$V_P=V^{\text{S}}_{P'\leftrightarrow P}+V^{\text{F}}_P$. 
The $V^{\text{S}}_{P'\leftrightarrow P}$ notations indicates that the slow
part of the potential is evaluated on a ``contracted'' ring polymer. 
The $P$-bead path can be Fourier-interpolated down to a smaller  number
of replicas P', for which energy and forces are computed, and then 
Fourier-expanded back to $P$ replicas. 
This makes RPC fully transparent from the point
of view of the integration of the equations of motion and the evaluation
of system properties, simplifying both notation and implementation.
Note that it is often the case that one wants to use a low
level of theory for the quickly varying part, $V^{\text{cheap}}\equiv V^{\text{F}}$,
and apply on a longer
time step or contracted ring polymer a correction term based
on a more accurate assessment of the inter-atomic potential 
$V^{\text{accu}} -V^{\text{cheap}} \equiv V^{\text{S}}$.

\begin{figure}
\centering
\includegraphics[width=1.0\columnwidth]{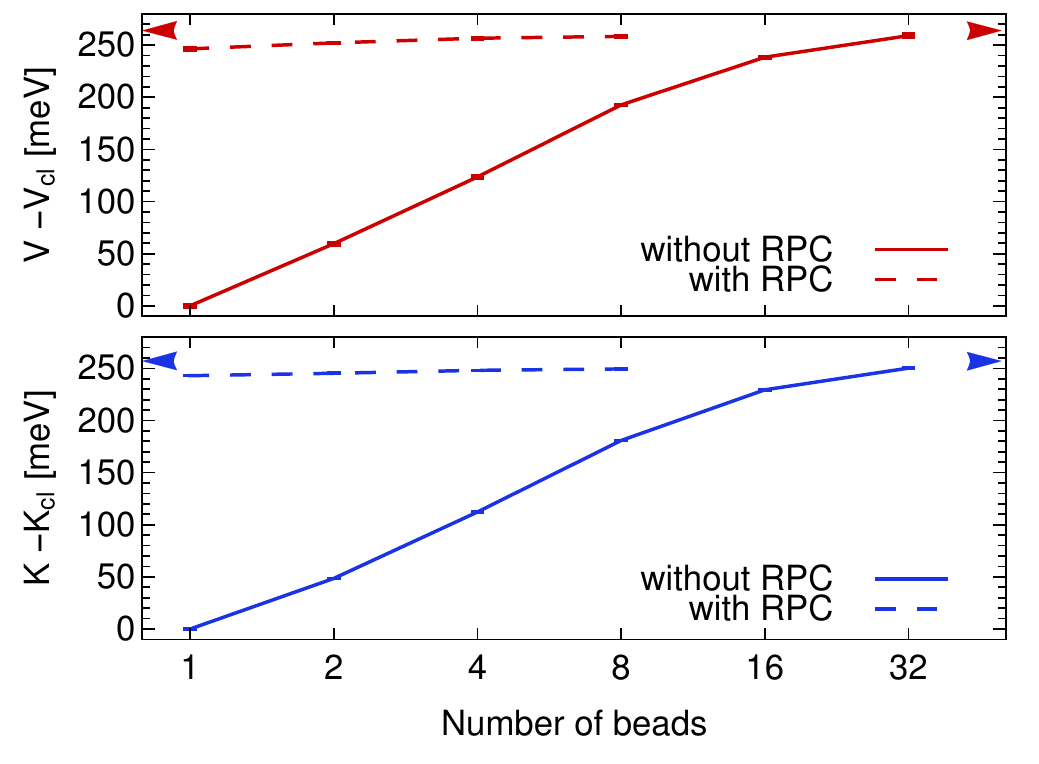}
\caption{Difference between quantum and  classical expectation values of potential (top) and kinetic energy (bottom) per molecule as a function of the number of path integral replicas for a simulation of q-TIP4P/F
liquid water at room temperature. Solid curves correspond
to full PIMD, while dashed curves correspond to a RPC setup
in which the bonded (stretch and bend) terms were computed
on 32 replicas, and the non-bonded 
(electrostatic and dispersion) terms were computed on a reduced number 
of replicas, that corresponds to the horizontal axis. The blue and red arrow-heads correspond to full PIMD using 64 beads.}  
\label{fig:h2o-rpc}
\end{figure}

As shown in Fig.~\ref{fig:h2o-rpc}, the convergence of observables
to quantum expectation values in a simulation of liquid water at room
temperature require a large number of replicas, of the order of $P \approx 30$. 
However, this is mostly due to the internal modes of each water molecule,
that are well described by cheap quasi-harmonic terms. The non-bonded 
dispersion and electrostatic interactions vary on considerably larger
length scales, and therefore a RPC scheme that computes the latter on a 
reduced number of replicas converges very rapidly. A simple
range-separation procedure would allow one to 
reach convergence with a single
evaluation of electrostatic interactions~\cite{mark-mano08cpl}. 
Here we want however just to verify to what extent one can push a 
MTS procedure, alone or in combination with RPC, in a PIMD context.

A multiple time-step integrator for constant-temperature PIMD
is epitomized by the corresponding split  operator
for the propagator over the outer time step $\Delta t$:
\begin{equation}
\begin{split}
 e^{-\mathcal{L}\Delta t}= &  e^{-(\mathcal{L}_\gamma + \mathcal{L}_0 + \mathcal{L}_{V}^{\text{F}} + \mathcal{L}_{V}^{\text{S}})\Delta t}
\approx
e^{-\mathcal{L}_\gamma\frac{\Delta t}{2}}
e^{-\mathcal{L}_{V}^{\text{S}}\frac{\Delta t}{2}} \times \\
\times &
\left[e^{-\mathcal{L}_{V}^{\text{F}}\frac{\Delta t}{2 M}} e^{-\mathcal{L}_0\frac{\Delta t}{M}} e^{-\mathcal{L}_{V}^{\text{F}}\frac{\Delta t}{2 M}}\right]^M
e^{-\mathcal{L}_{V}^{\text{S}}\frac{\Delta t}{2}}e^{-\mathcal{L}_\gamma\frac{\Delta t}{2}}.
\end{split}
\label{eq:liouville}
\end{equation}
First, a thermostat is applied by half the outer time step
($\mathcal{L}_\gamma \frac{\Delta t}{2}$); then the slowly varying
force (possibly computed by RPC) is applied 
($\mathcal{L}_V^\text{S} \frac{\Delta t}{2}$); then, 
the inner loop of the MTS integrator is performed, repeating
$P'$ times the combined integration of the 
fast force component ($\mathcal{L}_V^\text{F}$)
and the free-ring-polymer Hamiltonian 
($\mathcal{L}_0$), with a time step duly scaled by a factor of
$M$. Finally, the slowly varying force and the thermostat
are applied symmetrically, concluding the time step. 
As discussed e.g. in Refs.\cite{Ma-jsc51-verlet_impulse_instability,barth_jcp98_mts_extra_vs_imp}, regardless of how
well separated the two components of the potential
are, it is not possible to extend indefinitely 
the outer time step, as accumulation of errors 
will make the integration unstable when $\Delta t$
is larger than a fraction of the highest-frequency 
vibrational period. 

\begin{figure}
\centering
\includegraphics[width=1.0\columnwidth]{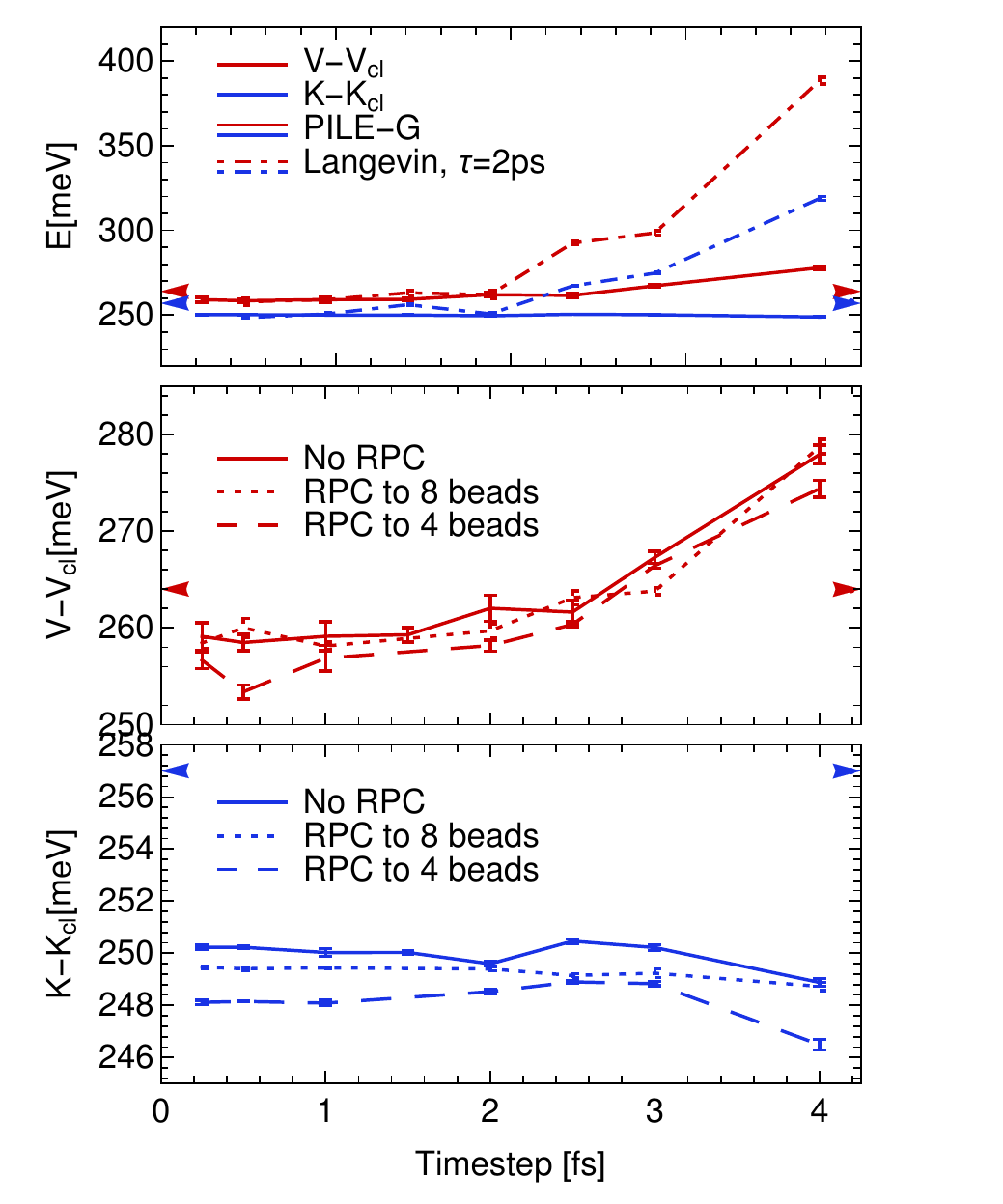}
\caption{Difference between quantum and classical expectation values of potential (top and middle)
 and kinetic (top and bottom) energy per molecule for a simulation of q-TIP4P/F liquid water at 
 room temperature, as a function of the outer time step. The bonded terms were computed with a 
 fixed inner time step of 0.25~fs, and propagated just outside the free ring-polymer part of the propagator. 
 The non-bonded terms were computed less often, with a
time interval as indicated on the horizontal axis. The blue and red curves correspond, respectively,
 to kinetic and potential energy. Solid curves correspond to 32 beads full PIMD simulation while dotted 
 and dashed curves to simulations in which the non-bonded terms were contracted to $4$ and $8$ beads 
 respectively. The dot dashed curves in the top panel correspond to simulations using a white noise
Langevin thermostat with a time constant of $2$ ps while the rest to simulations using
 a PILE-G thermostat.}
\label{fig:h2o-mts}
\end{figure}

Due to the presence of high-frequency
normal modes in the free ring-polymer Hamiltonian~(Eq. \eqref{eq:free-rp})
one would expect the first occurrence of resonances
to appear already at a time step of about 1fs for a path integral 
simulation with 32 replicas at 300K\cite{lueh+14jcp},
much earlier than the $\approx$2.7fs limit that would
be expected due to the stretching mode frequency.
Figure~\ref{fig:h2o-mts} shows that in this 
particular case the PIMD simulation remains stable
up to an outer step of about 2fs, even in a 
weakly-thermostatted (white-noise Langevin, thermostat correlation
time of 2 ps) PIMD simulation.
This rather puzzling finding can probably be ascribed to 
the exceedingly weak coupling between the physical potential
and the high-frequency normal modes of the ring polymer, 
and does not necessarily apply to different systems with 
stronger anharmonicities.
Here, increasing the dynamical masses of high-frequency 
ring-polymer modes would delay only minimally 
the resonance barrier, and would not
be possible when using approximate quantum dynamics 
techniques, such as centroid molecular dynamics 
(CMD)~\cite{cao-voth94jcp} and 
thermostatted ring polymer molecular dynamics
(TRPMD)~\cite{crai-mano04jcp,ross+14jcp}, that involve
specific prescriptions for the magnitude of the dynamical masses.
However, a similarly effective solution to deal with ring-polymer 
resonance barriers is to selectively thermalize the high-frequency vibrations~\cite{morr+11jcp}.
Exploiting the approximate knowledge of high-frequency
ring polymer modes, one can here simply use
optimally-damped Langevin thermostats 
in the (ring-polymer) normal-modes representation, as 
explained e.g. in Ref.~\cite{ceri+10jcp}.
As discussed in Ref.~\cite{ross+14jcp2}, however, some 
care must be paid to minimize the effect of normal modes
thermostatting on the dynamical properties of interest.

Figure~\ref{fig:h2o-mts} shows that the use of an optimally-damped
path integral Langevin thermostat~\cite{ceri+10jcp} does actually
extend the stability of the method to outer timesteps well above
the resonance barrier for the stretches.
In fact, the deviations that are observed for the mean 
potential energy are due to the outer time step becoming 
inappropriate to integrate the long-range force (which is relatively
quickly varying since we do not use any kind of range separation
scheme). This additional stabilization is probably due
to the coupling between ring-polymer modes and high-frequency
centroid vibrations -- which is also the cause for spectral
broadening observed in thermostatted RPMD.

\begin{figure}
\centering
\includegraphics[width=1\columnwidth]{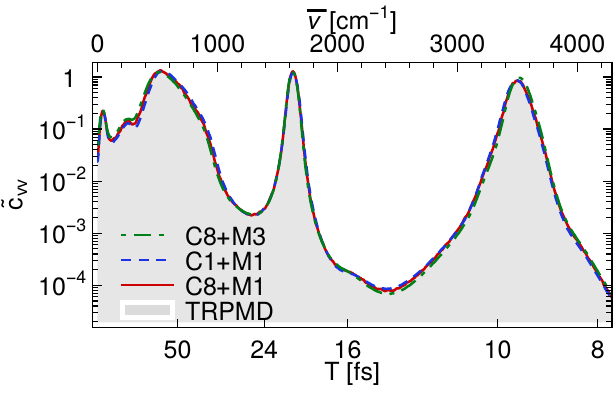}
\caption{The velocity-velocity autocorrelation function from 
TRPMD\cite{ross+14jcp} ($\lambda=0.5$) runs for
q-TIP4P/F water at 300K.
A reference run (timestep 0.25fs, 32 beads, shaded area)
is compared with a run with 8-beads RPC for the non-bonded
interactions, and an outer MTS timestep of 1fs (C8+M1, red 
continuous line), with a run with 1-bead RPC and 1fs MTS (C1+M1, blue dashed line), and with a run with 8-beads RPC and 3fs MTS (C8+M3, green dot-dashed line). 
 } 
\label{fig:acf}
\end{figure}

Figure~\ref{fig:acf} shows the velocity-velocity autocorrelation 
function for liquid water computed with TRPMD, and compare the 
results using  full 32-beads PIMD and those obtained 
with RPC to 8 beads of the non-bonded interactions, and MTS 
to 1fs outer. Dynamical properties appear to be less strongly
affected than thermodynamic averages when RPC/MTS techniques
are pushed to their limits. Using RPC down to the centroid, or using a 
3fs outer time step, lead to minor deviations from the reference.
In summary, these benchmarks demonstrate the effectiveness of our
implementation of combined RPC and MTS techniques in i-PI, 
and shows that with an appropriate thermostatting scheme one
can push the stability limit of MTS schemes beyond the 
resonance barrier of RPMD, for the calculation of both
static and dynamical properties (Fig.~\ref{fig:h2o-mts} and~\ref{fig:acf}).

\section{Nuclear quantum effects and electronic structure methods: the zundel cation}

Having briefly reviewed the main ideas, and demonstrated the stability
of RPC+MTS PIMD, let us move on to discuss a few aspects that are
specific to a first principles context. 
Here, we simulate the quantum distributions of the Zundel 
cation at the MP2 level of theory,
using semi-local DFT for RPC+MTS. While other choices are 
possible, this combination is really attractive, as it is 
general in nature, and the impact of the scheme 
can be very significant. Indeed, as system size increases, 
MP2 and semi-local DFT display very different computational costs, 
since traditional implementations scale as $O(N^5)$ and $O(N^3)$ respectively. 
For example, three orders of magnitude time difference 
can be expected for samples of 64 water molecules~\cite{delb+15jcp2}.
The excellent scalability of our MP2 implementation\cite{delb+13jctc,delb+15cpc} 
nevertheless enables an acceptable time to solution.
In the MP2+RPC context a large number of cheap DFT calculations 
must be combined with the expensive MP2 calculations,
and the question of load balancing,
avoiding idle processes as much as possible, imposes itself. 
As a simple, yet efficient, solution to this problem, 
we over-subscribe compute nodes using two processes per core.
One process belongs to the $P$ DFT tasks, while
one process belongs to the $P'$ MP2 tasks and the contracted
DFT tasks, which run at the same time as MP2\footnote{
To achieve an effective load balancing, if $Q$ is the number
of processes assigned to each of the $P$ DFT tasks, and
$C_\text{MP2}$ and $C_\text{DFT}$ are the scalar costs of 
each force evaluation at the two levels of theory,
each of the contracted MP2 processes should be assigned
about $P Q C_\text{MP2}/(m(C_\text{MP2}+C_\text{DFT}))$ 
processes, and 
$P Q C_\text{DFT}/(m(C_\text{MP2}+C_\text{DFT}))$ processes should be
given to each contracted DFT task.
}. 
While the MP2 tasks computes, the full-path DFT tasks are 
effectively sleeping, and vice versa. In this way, idle 
resources are avoided, and only a small fraction of total 
wall time is spent in the DFT part, which is naturally 
parallel over the beads, while the computation for each 
bead is parallelised as well. If the contraction is pushed
all the way to the centroid, the cost of 
performing MP2+DFT with MTS+RPC is basically the same  as the cost of 
doing standard MP2 MD. An example of how these simulations
can be set up in practice can be obtained from the authors.

\begin{table}[hpbt]
\centering
\begin{tabular}{l c c c}
\hline\hline
Simulation & $K_\text{H}$ [meV] & $K_\text{O}$ [meV] & $d_\text{OO}$ [\AA] \\
\hline                                                  
 MP2       &   148.4(5)    &   58.9(4)  &   2.413(4)  \\
 PBE       &   143.1(5)    &   57.5(4)  &   2.445(4)  \\
 PBE+C     &   143.2(5)    &   58.5(3)  &   2.409(5)  \\ 
 PBE+C2+M  &   144.7(6)    &   58.3(4)  &   2.410(1)  \\
 ODF       &   148.9(5)    &   58.9(4)  &   2.416(3)  \\
 ODF+C     &   149.3(5)    &   59.6(4)  &   2.413(4)  \\
 ODF+C+M   &   149.7(3)    &   59.3(3)  &   2.410(3)  \\
 ODF+C+M2  &   149.4(4)    &   59.4(3)  &   2.412(5)  \\
 ODF+C+M+G &   148.0(9)    &   59.4(5)  &   2.412(2)  \\
 HFX+C+M   &   155.3(7)    &   60.6(5)  &   2.411(3)  \\
 \hline\hline
\end{tabular}
\caption{Expectation values of quantum kinetic energies per
H and O atom, and for the O-O distance, in a simulation of a gas-phase Zundel cation at 300K. All simulations were 
performed using 32 beads and a base time step of $0.25$fs, except that using PIGLET (+G) that
used six replicas. PBE refers to use of the standard PBE generalized gradient approximation (GGA)
functional~\cite{perd+96prl}, ODF refers to a GGA optimized to MP2\cite{delb+15jcp2},
and HFX to Hartree-Fock. 
Ring-polymer contraction was used
to reduce the cost by computing the MP2 forces on the centroid 
only (+C), or on two beads (+C2). 
A multiple time step algorithm was also employed to evaluate such force 
only once every 1fs (+M) or 2fs (+M2). 
All simulations were run for 10ps, including 1ps for equilibration. 
Statistical errors on
the last digit are reported in parentheses. }
\label{tab:zundel}
\end{table}

The combination of MP2 with semi-local DFT 
employs the same computational setup as described 
in Ref.~\cite{delb+15jcp2}. In particular, this 
approach is based on the resolution of identity Gaussian and 
Plane Waves (RI-GPW) which provides an efficient and 
scalable approach to perform MP2 based MD in gas and 
condensed phases.\cite{delb+12jctc,delb+13jctc,delb+15cpc,delb+15jcp} 
The Gaussian basis employed for MP2 is of the correlation 
consistent triple zeta quality\cite{delb+12jctc} and
is parametrized for the pseudopotentials 
employed\cite{goed-physrevb-sep_sdsgpp}
The optimized density functional (ODF) discussed below is of the GGA family, starting from the PBE1W 
\cite{dahlk-jpcb-improved_functionals_h2o} functional,
for which the small basis and van der Waals D3 
parameters\cite{grimme-jcp10-dftd_disp_functional} have been specifically
refitted in order to reproduce the energetics of bulk liquid water.\cite{delb+15jcp2}

\begin{figure}
\centering
\includegraphics[width=1\columnwidth]{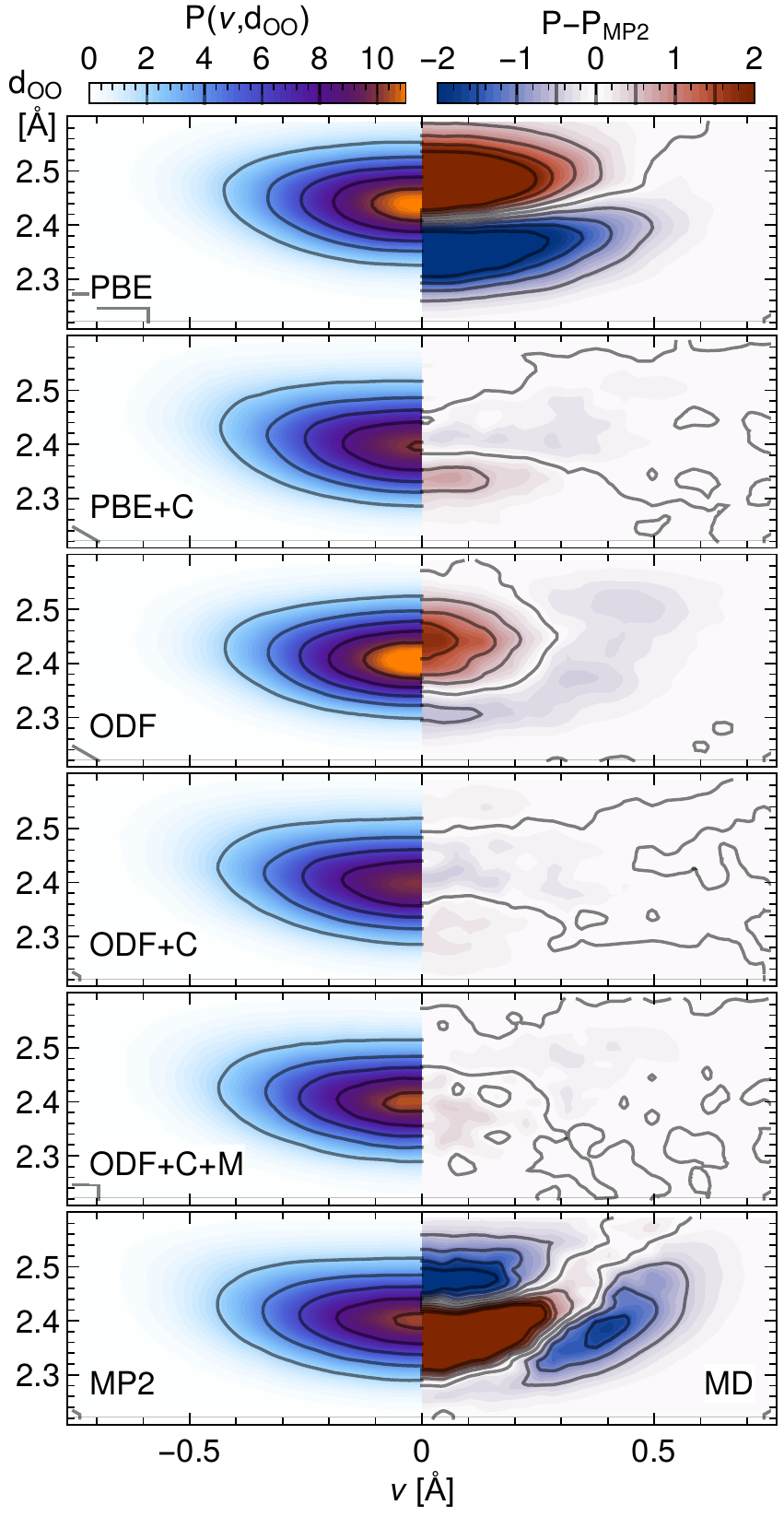}
\caption{Joint probability distribution for the proton transfer coordinate and 
the O-O distance in the gas-phase Zundel cation at 300K, modeled with
nuclear quantum effects and different electronic structure methods (left panels). 
Right-hand panels show the difference with respect to the MP2 reference.
The bottom-right panel shows the difference between a classical MP2 MD simulation
and the (quantum) MP2 reference. } 
\label{fig:doonu}
\end{figure}

Results for the mean kinetic energy of hydrogen and oxygen atoms, 
and the mean O-O distance, for different simulations, are reported
in Table~\ref{tab:zundel}. The table caption summarizes the details 
of the simulations. The standard PBE functional underestimates the H kinetic energy
with single-bead contraction, improving as expected with a contraction to two beads. 
This indicates that PBE differs too much from the reference, 
requiring a considerable increase in the number of MP2 replicas. Also Hartree-Fock is not very accurate,
this time predicting a too large value of the H kinetic energy, despite the RPC correction.
This suggests that the MTS strategy proposed in Ref.~\cite{stee13jcp}, which combines HFX and MP2, cannot 
be used universally for RPC, at least not without a large number of 
correction beads. Hybrid functionals, which mix GGA exchange with Hartree-Fock exchange, 
will likely provide a suitable intermediate potential energy surface.
However, as a computationally less demanding, but less universal alternative,
an optimized semi-local DFT (ODF) shows excellent performance.
The results reported in Table~\ref{tab:zundel} are within the statistical error bars 
from the MP2 reference already without any RPC correction, 
and as a consequence it is not possible to detect statistically significant
effects on these averages due to RPC and/or MTS. 
However, also in this case the RPC correction does have a noticeable effect.
This is shown in Figure~\ref{fig:doonu} using the joint probability distribution 
of the O-O distance and the proton transfer coordinate. The difference between 
the MP2 and the ODF+C is almost zero and smaller than ODF only. 
The remaining error is largely due to the statistical uncertainty in the 
probability distributions, which is relatively costly to reduce for the 
reference MP2 simulation that employs 32 MP2 beads.

Let us finally note that RPC and MTS methods can be seamlessly 
combined with other strategies to reduce the cost of a PIMD calculation:
as demonstrated in Table~\ref{tab:zundel}, the use of a colored-noise
PIGLET thermostat~\cite{ceri-mano12prl} makes it possible to reduce
the number of baseline beads to six, that would be advantageous
in cases where the cost of the GGA calculations is not negligible.

\begin{figure}
\centering
\includegraphics[width=1\columnwidth]{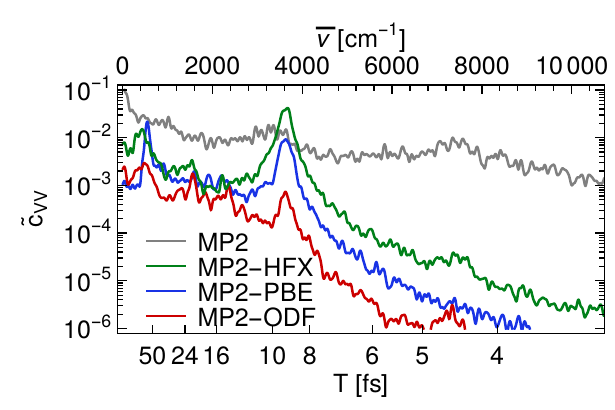}
\caption{Fourier-transform of the potential-potential correlation 
function for different RPC simulations. The curve for MP2 (in grey)
is taken as a reference; being evaluated on 32 replicas, it also 
contains the high-frequency non-centroid modes of the ring polymer.
The other two curves correspond to the correction potential
(MP2 minus GGA) evaluated on the centroid for the HFX (green), PBE (blue)
and optimized DFT (red) simulations. The horizontal scale indicates
the period of different vibrational modes, for ease of reference.
Since these simulations were heavily thermostatted for sampling
and stability, the spectrum has no physical meaning, and is just
a tool to assess the time scales that are relevant for different
potential components.
} 
\label{fig:cvv}
\end{figure}

A simple way to assess the viability of accelerating a simulation
by means of RPC and/or MTS is to verify the magnitude and time scale
for the variation of the correction potential. Fig.~\ref{fig:cvv}
shows the Fourier transform of the correction potential
correlation function $\left<V^\text{S}(t)V^\text{S}(0)\right>$.
Clearly, the difference between MP2 and HFX contains strong 
high-frequency components, which would require a high number of
beads in the contracted $V^\text{S}$. Using the PBE GGA shows only 
a marginal advantage, as the spectrum still has a pronounced peak
at the stretching frequencies. On the contrary, the 
ODF exhibits a smaller difference, and in particular
a density of states smaller by an order of magnitude for the 
stretching region - which explains the excellent performance
in reproducing quantum kinetic energies even without RPC.

\section{Conclusions}

Even on a small model system such as the gas-phase zundel cation
a combination of ring-polymer contraction, multiple time step
algorithms and a force-matching strategy to minimise the discrepancy
between the expensive, post-HF calculation and a GGA DFT baseline
makes it possible to model nuclear quantum effects at virtually 
no additional cost. Modular implementation in i-PI means that all
sorts of similar schemes -- combining different levels of theory,
more or less converged basis sets, range-separated potentials, 
empirical or semi-empirical methods, etc. -- could be tested and 
used without further coding effort\footnote{The development version
of i-PI can be obtained from the Authors.}. Parallelism across replicas and 
potential components can be obtained by appropriately managing
the execution on a HPC system, realizing a rudimentary but effective
form of load balancing. Multiple time-stepping and ring-polymer contraction
provide additional tools to reduce the computational
expense for modeling quantum nuclei in ab initio
simulations, that can be combined seamlessly with 
with high-order path integrals and correlated-noise
techniques.
Nuclear quantum effects can now be incorporated accurately and 
at low cost in simulations employing high level electronic 
structure theory. Given the importance of nuclear quantum effects 
and the generality of the approach, we expect this to become 
standard practice in the near future.

\section{Acknowledgements}
J.V. acknowledges financial support by the European Union
FP7 in the form of an ERC Starting Grant under contract no. 277910.
M.C and V.K acknowledge financial support by the Swiss National
Science Foundation (project ID 200021-159896).
This research was partly supported by NCCR MARVEL, funded by the 
Swiss National Science Foundation.
Calculations were enabled by the Swiss National Supercomputer 
Centre (CSCS) under project ID ch5.
During the preparation of this manuscript we became aware that
O. Marsalek and T. Markland were working on a related 
combination of DFT and self-consistent tight binding~\cite{TE-PC} using 
MTS and RPC -- another excellent example of how inexpensive modelling
of quantum nuclei can be achieved by coupling different levels
of electronic structure theory.


\begin{thebibliography}{47}%
\makeatletter
\providecommand \@ifxundefined [1]{%
 \@ifx{#1\undefined}
}%
\providecommand \@ifnum [1]{%
 \ifnum #1\expandafter \@firstoftwo
 \else \expandafter \@secondoftwo
 \fi
}%
\providecommand \@ifx [1]{%
 \ifx #1\expandafter \@firstoftwo
 \else \expandafter \@secondoftwo
 \fi
}%
\providecommand \natexlab [1]{#1}%
\providecommand \enquote  [1]{``#1''}%
\providecommand \bibnamefont  [1]{#1}%
\providecommand \bibfnamefont [1]{#1}%
\providecommand \citenamefont [1]{#1}%
\providecommand \href@noop [0]{\@secondoftwo}%
\providecommand \href [0]{\begingroup \@sanitize@url \@href}%
\providecommand \@href[1]{\@@startlink{#1}\@@href}%
\providecommand \@@href[1]{\endgroup#1\@@endlink}%
\providecommand \@sanitize@url [0]{\catcode `\\12\catcode `\$12\catcode
  `\&12\catcode `\#12\catcode `\^12\catcode `\_12\catcode `\%12\relax}%
\providecommand \@@startlink[1]{}%
\providecommand \@@endlink[0]{}%
\providecommand \url  [0]{\begingroup\@sanitize@url \@url }%
\providecommand \@url [1]{\endgroup\@href {#1}{\urlprefix }}%
\providecommand \urlprefix  [0]{URL }%
\providecommand \Eprint [0]{\href }%
\providecommand \doibase [0]{http://dx.doi.org/}%
\providecommand \selectlanguage [0]{\@gobble}%
\providecommand \bibinfo  [0]{\@secondoftwo}%
\providecommand \bibfield  [0]{\@secondoftwo}%
\providecommand \translation [1]{[#1]}%
\providecommand \BibitemOpen [0]{}%
\providecommand \bibitemStop [0]{}%
\providecommand \bibitemNoStop [0]{.\EOS\space}%
\providecommand \EOS [0]{\spacefactor3000\relax}%
\providecommand \BibitemShut  [1]{\csname bibitem#1\endcsname}%
\let\auto@bib@innerbib\@empty
%</preamble>
\bibitem [{\citenamefont {{Del Ben}}\ \emph {et~al.}(2013)\citenamefont {{Del
  Ben}}, \citenamefont {Sch{\"{o}}nherr}, \citenamefont {Hutter},\ and\
  \citenamefont {Vandevondele}}]{delb+13jpcl}%
  \BibitemOpen
  \bibfield  {author} {\bibinfo {author} {\bibfnamefont {M.}~\bibnamefont {{Del
  Ben}}}, \bibinfo {author} {\bibfnamefont {M.}~\bibnamefont
  {Sch{\"{o}}nherr}}, \bibinfo {author} {\bibfnamefont {J.}~\bibnamefont
  {Hutter}}, \ and\ \bibinfo {author} {\bibfnamefont {J.}~\bibnamefont
  {Vandevondele}},\ }\href@noop {} {\bibfield  {journal} {\bibinfo  {journal}
  {J. Phys. Chem. Lett.}\ }\textbf {\bibinfo {volume} {4}},\ \bibinfo {pages}
  {3753} (\bibinfo {year} {2013})}\BibitemShut {NoStop}%
\bibitem [{\citenamefont {{Del Ben}}, \citenamefont {Hutter},\ and\
  \citenamefont {VandeVondele}(2015{\natexlab{a}})}]{delb+15jcp2}%
  \BibitemOpen
  \bibfield  {author} {\bibinfo {author} {\bibfnamefont {M.}~\bibnamefont {{Del
  Ben}}}, \bibinfo {author} {\bibfnamefont {J.}~\bibnamefont {Hutter}}, \ and\
  \bibinfo {author} {\bibfnamefont {J.}~\bibnamefont {VandeVondele}},\
  }\href@noop {} {\bibfield  {journal} {\bibinfo  {journal} {J. Chem. Phys.}\
  }\textbf {\bibinfo {volume} {143}},\ \bibinfo {pages} {054506} (\bibinfo
  {year} {2015}{\natexlab{a}})}\BibitemShut {NoStop}%
\bibitem [{\citenamefont {Zen}\ \emph {et~al.}(2015)\citenamefont {Zen},
  \citenamefont {Luo}, \citenamefont {Mazzola}, \citenamefont {Guidoni},\ and\
  \citenamefont {Sorella}}]{zen+15jcp}%
  \BibitemOpen
  \bibfield  {author} {\bibinfo {author} {\bibfnamefont {A.}~\bibnamefont
  {Zen}}, \bibinfo {author} {\bibfnamefont {Y.}~\bibnamefont {Luo}}, \bibinfo
  {author} {\bibfnamefont {G.}~\bibnamefont {Mazzola}}, \bibinfo {author}
  {\bibfnamefont {L.}~\bibnamefont {Guidoni}}, \ and\ \bibinfo {author}
  {\bibfnamefont {S.}~\bibnamefont {Sorella}},\ }\href@noop {} {\bibfield
  {journal} {\bibinfo  {journal} {J. Chem. Phys.}\ }\textbf {\bibinfo {volume}
  {142}},\ \bibinfo {pages} {144111} (\bibinfo {year} {2015})}\BibitemShut
  {NoStop}%
\bibitem [{\citenamefont {Feynman}\ and\ \citenamefont
  {Hibbs}(1964)}]{feyn-hibb65book}%
  \BibitemOpen
  \bibfield  {author} {\bibinfo {author} {\bibfnamefont {R.~P.}\ \bibnamefont
  {Feynman}}\ and\ \bibinfo {author} {\bibfnamefont {A.~R.}\ \bibnamefont
  {Hibbs}},\ }\href@noop {} {\emph {\bibinfo {title} {{Quantum Mechanics and
  Path Integrals}}}}\ (\bibinfo  {publisher} {McGraw-Hill},\ \bibinfo {address}
  {New York},\ \bibinfo {year} {1964})\BibitemShut {NoStop}%
\bibitem [{\citenamefont {Chandler}\ and\ \citenamefont
  {Wolynes}(1981)}]{chan-woly81jcp}%
  \BibitemOpen
  \bibfield  {author} {\bibinfo {author} {\bibfnamefont {D.}~\bibnamefont
  {Chandler}}\ and\ \bibinfo {author} {\bibfnamefont {P.~G.}\ \bibnamefont
  {Wolynes}},\ }\href@noop {} {\bibfield  {journal} {\bibinfo  {journal} {J.
  Chem. Phys.}\ }\textbf {\bibinfo {volume} {74}},\ \bibinfo {pages} {4078}
  (\bibinfo {year} {1981})}\BibitemShut {NoStop}%
\bibitem [{\citenamefont {Parrinello}\ and\ \citenamefont
  {Rahman}(1984)}]{parr-rahm84jcp}%
  \BibitemOpen
  \bibfield  {author} {\bibinfo {author} {\bibfnamefont {M.}~\bibnamefont
  {Parrinello}}\ and\ \bibinfo {author} {\bibfnamefont {A.}~\bibnamefont
  {Rahman}},\ }\href@noop {} {\bibfield  {journal} {\bibinfo  {journal} {J.
  Chem. Phys.}\ }\textbf {\bibinfo {volume} {80}},\ \bibinfo {pages} {860}
  (\bibinfo {year} {1984})}\BibitemShut {NoStop}%
\bibitem [{\citenamefont {Jang}\ and\ \citenamefont
  {Voth}(2001)}]{jang-voth01jcp}%
  \BibitemOpen
  \bibfield  {author} {\bibinfo {author} {\bibfnamefont {S.~S.}\ \bibnamefont
  {Jang}}\ and\ \bibinfo {author} {\bibfnamefont {G.~A.}\ \bibnamefont
  {Voth}},\ }\href@noop {} {\bibfield  {journal} {\bibinfo  {journal} {J. Chem.
  Phys.}\ }\textbf {\bibinfo {volume} {115}},\ \bibinfo {pages} {7832}
  (\bibinfo {year} {2001})}\BibitemShut {NoStop}%
\bibitem [{\citenamefont {P{\'{e}}rez}\ and\ \citenamefont
  {Tuckerman}(2011)}]{pere-tuck11jcp}%
  \BibitemOpen
  \bibfield  {author} {\bibinfo {author} {\bibfnamefont {A.}~\bibnamefont
  {P{\'{e}}rez}}\ and\ \bibinfo {author} {\bibfnamefont {M.~E.}\ \bibnamefont
  {Tuckerman}},\ }\href@noop {} {\bibfield  {journal} {\bibinfo  {journal} {J.
  Chem. Phys.}\ }\textbf {\bibinfo {volume} {135}},\ \bibinfo {pages} {064104}
  (\bibinfo {year} {2011})}\BibitemShut {NoStop}%
\bibitem [{\citenamefont {Ceriotti}, \citenamefont {Bussi},\ and\ \citenamefont
  {Parrinello}(2009)}]{ceri+09prl2}%
  \BibitemOpen
  \bibfield  {author} {\bibinfo {author} {\bibfnamefont {M.}~\bibnamefont
  {Ceriotti}}, \bibinfo {author} {\bibfnamefont {G.}~\bibnamefont {Bussi}}, \
  and\ \bibinfo {author} {\bibfnamefont {M.}~\bibnamefont {Parrinello}},\
  }\href@noop {} {\bibfield  {journal} {\bibinfo  {journal} {Phys. Rev. Lett.}\
  }\textbf {\bibinfo {volume} {103}},\ \bibinfo {pages} {30603} (\bibinfo
  {year} {2009})}\BibitemShut {NoStop}%
\bibitem [{\citenamefont {Ceriotti}, \citenamefont {Manolopoulos},\ and\
  \citenamefont {Parrinello}(2011)}]{ceri+11jcp}%
  \BibitemOpen
  \bibfield  {author} {\bibinfo {author} {\bibfnamefont {M.}~\bibnamefont
  {Ceriotti}}, \bibinfo {author} {\bibfnamefont {D.~E.}\ \bibnamefont
  {Manolopoulos}}, \ and\ \bibinfo {author} {\bibfnamefont {M.}~\bibnamefont
  {Parrinello}},\ }\href@noop {} {\bibfield  {journal} {\bibinfo  {journal} {J.
  Chem. Phys.}\ }\textbf {\bibinfo {volume} {134}},\ \bibinfo {pages} {84104}
  (\bibinfo {year} {2011})}\BibitemShut {NoStop}%
\bibitem [{\citenamefont {Ceriotti}\ and\ \citenamefont
  {Manolopoulos}(2012)}]{ceri-mano12prl}%
  \BibitemOpen
  \bibfield  {author} {\bibinfo {author} {\bibfnamefont {M.}~\bibnamefont
  {Ceriotti}}\ and\ \bibinfo {author} {\bibfnamefont {D.~E.}\ \bibnamefont
  {Manolopoulos}},\ }\href@noop {} {\bibfield  {journal} {\bibinfo  {journal}
  {Phys. Rev. Lett.}\ }\textbf {\bibinfo {volume} {109}},\ \bibinfo {pages}
  {100604} (\bibinfo {year} {2012})}\BibitemShut {NoStop}%
\bibitem [{\citenamefont {Poma}\ and\ \citenamefont {{Delle
  Site}}(2010)}]{poma-dell10prl}%
  \BibitemOpen
  \bibfield  {author} {\bibinfo {author} {\bibfnamefont {A.~B.}\ \bibnamefont
  {Poma}}\ and\ \bibinfo {author} {\bibfnamefont {L.}~\bibnamefont {{Delle
  Site}}},\ }\href@noop {} {\bibfield  {journal} {\bibinfo  {journal} {Phys.
  Rev. Lett.}\ }\textbf {\bibinfo {volume} {104}},\ \bibinfo {pages} {250201}
  (\bibinfo {year} {2010})}\BibitemShut {NoStop}%
\bibitem [{\citenamefont {Markland}\ and\ \citenamefont
  {Manolopoulos}(2008{\natexlab{a}})}]{mark-mano08jcp}%
  \BibitemOpen
  \bibfield  {author} {\bibinfo {author} {\bibfnamefont {T.~E.}\ \bibnamefont
  {Markland}}\ and\ \bibinfo {author} {\bibfnamefont {D.~E.}\ \bibnamefont
  {Manolopoulos}},\ }\href@noop {} {\bibfield  {journal} {\bibinfo  {journal}
  {J. Chem. Phys.}\ }\textbf {\bibinfo {volume} {129}},\ \bibinfo {pages}
  {024105} (\bibinfo {year} {2008}{\natexlab{a}})}\BibitemShut {NoStop}%
\bibitem [{\citenamefont {Markland}\ and\ \citenamefont
  {Manolopoulos}(2008{\natexlab{b}})}]{mark-mano08cpl}%
  \BibitemOpen
  \bibfield  {author} {\bibinfo {author} {\bibfnamefont {T.~E.}\ \bibnamefont
  {Markland}}\ and\ \bibinfo {author} {\bibfnamefont {D.~E.}\ \bibnamefont
  {Manolopoulos}},\ }\href@noop {} {\bibfield  {journal} {\bibinfo  {journal}
  {Chem. Phys. Lett.}\ }\textbf {\bibinfo {volume} {464}},\ \bibinfo {pages}
  {256} (\bibinfo {year} {2008}{\natexlab{b}})}\BibitemShut {NoStop}%
\bibitem [{\citenamefont {Martyna}, \citenamefont {Tuckerman},\ and\
  \citenamefont {Klein}(1992)}]{mart+92jcp}%
  \BibitemOpen
  \bibfield  {author} {\bibinfo {author} {\bibfnamefont {G.~J.}\ \bibnamefont
  {Martyna}}, \bibinfo {author} {\bibfnamefont {M.~E.}\ \bibnamefont
  {Tuckerman}}, \ and\ \bibinfo {author} {\bibfnamefont {M.~L.}\ \bibnamefont
  {Klein}},\ }\href@noop {} {\bibfield  {journal} {\bibinfo  {journal} {J.
  Chem. Phys.}\ }\textbf {\bibinfo {volume} {97}},\ \bibinfo {pages} {2635}
  (\bibinfo {year} {1992})}\BibitemShut {NoStop}%
\bibitem [{\citenamefont {Zhou}\ \emph {et~al.}(2001)\citenamefont {Zhou},
  \citenamefont {Harder}, \citenamefont {Xu},\ and\ \citenamefont
  {Berne}}]{zhou+01jcp}%
  \BibitemOpen
  \bibfield  {author} {\bibinfo {author} {\bibfnamefont {R.}~\bibnamefont
  {Zhou}}, \bibinfo {author} {\bibfnamefont {E.}~\bibnamefont {Harder}},
  \bibinfo {author} {\bibfnamefont {H.}~\bibnamefont {Xu}}, \ and\ \bibinfo
  {author} {\bibfnamefont {B.~J.}\ \bibnamefont {Berne}},\ }\href@noop {}
  {\bibfield  {journal} {\bibinfo  {journal} {J. Chem. Phys.}\ }\textbf
  {\bibinfo {volume} {115}},\ \bibinfo {pages} {2348} (\bibinfo {year}
  {2001})}\BibitemShut {NoStop}%
\bibitem [{\citenamefont {Ferrarotti}\ \emph {et~al.}(2015)\citenamefont
  {Ferrarotti}, \citenamefont {Bottaro}, \citenamefont {P{\'{e}}rez-Villa},\
  and\ \citenamefont {Bussi}}]{ferr+15jctc}%
  \BibitemOpen
  \bibfield  {author} {\bibinfo {author} {\bibfnamefont {M.~J.}\ \bibnamefont
  {Ferrarotti}}, \bibinfo {author} {\bibfnamefont {S.}~\bibnamefont {Bottaro}},
  \bibinfo {author} {\bibfnamefont {A.}~\bibnamefont {P{\'{e}}rez-Villa}}, \
  and\ \bibinfo {author} {\bibfnamefont {G.}~\bibnamefont {Bussi}},\
  }\href@noop {} {\bibfield  {journal} {\bibinfo  {journal} {J. Chem. Theory
  Comput.}\ }\textbf {\bibinfo {volume} {11}},\ \bibinfo {pages} {139}
  (\bibinfo {year} {2015})}\BibitemShut {NoStop}%
\bibitem [{\citenamefont {Guidon}\ \emph {et~al.}(2008)\citenamefont {Guidon},
  \citenamefont {Schiffmann}, \citenamefont {Hutter},\ and\ \citenamefont
  {Vandevondele}}]{guid+08jcp}%
  \BibitemOpen
  \bibfield  {author} {\bibinfo {author} {\bibfnamefont {M.}~\bibnamefont
  {Guidon}}, \bibinfo {author} {\bibfnamefont {F.}~\bibnamefont {Schiffmann}},
  \bibinfo {author} {\bibfnamefont {J.}~\bibnamefont {Hutter}}, \ and\ \bibinfo
  {author} {\bibfnamefont {J.}~\bibnamefont {Vandevondele}},\ }\href@noop {}
  {\bibfield  {journal} {\bibinfo  {journal} {J. Chem. Phys.}\ }\textbf
  {\bibinfo {volume} {128}},\ \bibinfo {pages} {214104} (\bibinfo {year}
  {2008})}\BibitemShut {NoStop}%
\bibitem [{\citenamefont {VandeVondele}\ and\ \citenamefont
  {Rothlisberger}(2000)}]{vande-jcp00-free_calc}%
  \BibitemOpen
  \bibfield  {author} {\bibinfo {author} {\bibfnamefont {J.}~\bibnamefont
  {VandeVondele}}\ and\ \bibinfo {author} {\bibfnamefont {U.}~\bibnamefont
  {Rothlisberger}},\ }\href@noop {} {\bibfield  {journal} {\bibinfo  {journal}
  {The Journal of Chemical Physics}\ }\textbf {\bibinfo {volume} {113}},\
  \bibinfo {pages} {4863} (\bibinfo {year} {2000})}\BibitemShut {NoStop}%
\bibitem [{\citenamefont {Iftimie}\ \emph {et~al.}(2000)\citenamefont
  {Iftimie}, \citenamefont {Salahub}, \citenamefont {Wei},\ and\ \citenamefont
  {Schofield}}]{ifti-jcp00-class_pot}%
  \BibitemOpen
  \bibfield  {author} {\bibinfo {author} {\bibfnamefont {R.}~\bibnamefont
  {Iftimie}}, \bibinfo {author} {\bibfnamefont {D.}~\bibnamefont {Salahub}},
  \bibinfo {author} {\bibfnamefont {D.}~\bibnamefont {Wei}}, \ and\ \bibinfo
  {author} {\bibfnamefont {J.}~\bibnamefont {Schofield}},\ }\href@noop {}
  {\bibfield  {journal} {\bibinfo  {journal} {The Journal of Chemical Physics}\
  }\textbf {\bibinfo {volume} {113}},\ \bibinfo {pages} {4852} (\bibinfo {year}
  {2000})}\BibitemShut {NoStop}%
\bibitem [{\citenamefont {McGrath}\ \emph {et~al.}(2005)\citenamefont
  {McGrath}, \citenamefont {Siepmann}, \citenamefont {Kuo}, \citenamefont
  {Mundy}, \citenamefont {VandeVondele}, \citenamefont {Hutter}, \citenamefont
  {Mohamed},\ and\ \citenamefont {Krack}}]{mcgrath-cpc05-isobar_isoth_mc}%
  \BibitemOpen
  \bibfield  {author} {\bibinfo {author} {\bibfnamefont {M.~J.}\ \bibnamefont
  {McGrath}}, \bibinfo {author} {\bibfnamefont {J.~I.}\ \bibnamefont
  {Siepmann}}, \bibinfo {author} {\bibfnamefont {I.-F.~W.}\ \bibnamefont
  {Kuo}}, \bibinfo {author} {\bibfnamefont {C.~J.}\ \bibnamefont {Mundy}},
  \bibinfo {author} {\bibfnamefont {J.}~\bibnamefont {VandeVondele}}, \bibinfo
  {author} {\bibfnamefont {J.}~\bibnamefont {Hutter}}, \bibinfo {author}
  {\bibfnamefont {F.}~\bibnamefont {Mohamed}}, \ and\ \bibinfo {author}
  {\bibfnamefont {M.}~\bibnamefont {Krack}},\ }\href@noop {} {\bibfield
  {journal} {\bibinfo  {journal} {ChemPhysChem}\ }\textbf {\bibinfo {volume}
  {6}},\ \bibinfo {pages} {1894} (\bibinfo {year} {2005})}\BibitemShut
  {NoStop}%
\bibitem [{\citenamefont {Steele}(2013)}]{stee13jcp}%
  \BibitemOpen
  \bibfield  {author} {\bibinfo {author} {\bibfnamefont {R.~P.}\ \bibnamefont
  {Steele}},\ }\href@noop {} {\bibfield  {journal} {\bibinfo  {journal} {J.
  Chem. Phys.}\ }\textbf {\bibinfo {volume} {139}},\ \bibinfo {pages} {011102}
  (\bibinfo {year} {2013})}\BibitemShut {NoStop}%
\bibitem [{\citenamefont {Luehr}, \citenamefont {Markland},\ and\ \citenamefont
  {Mart{\'{\i}}nez}(2014)}]{lueh+14jcp}%
  \BibitemOpen
  \bibfield  {author} {\bibinfo {author} {\bibfnamefont {N.}~\bibnamefont
  {Luehr}}, \bibinfo {author} {\bibfnamefont {T.~E.}\ \bibnamefont {Markland}},
  \ and\ \bibinfo {author} {\bibfnamefont {T.~J.}\ \bibnamefont
  {Mart{\'{\i}}nez}},\ }\href@noop {} {\bibfield  {journal} {\bibinfo
  {journal} {J. Chem. Phys.}\ }\textbf {\bibinfo {volume} {140}},\ \bibinfo
  {pages} {084116} (\bibinfo {year} {2014})}\BibitemShut {NoStop}%
\bibitem [{\citenamefont {Fatehi}\ and\ \citenamefont
  {Steele}(2015)}]{fate-stee15jctc}%
  \BibitemOpen
  \bibfield  {author} {\bibinfo {author} {\bibfnamefont {S.}~\bibnamefont
  {Fatehi}}\ and\ \bibinfo {author} {\bibfnamefont {R.~P.}\ \bibnamefont
  {Steele}},\ }\href@noop {} {\bibfield  {journal} {\bibinfo  {journal} {J.
  Chem. Theory Comput.}\ }\textbf {\bibinfo {volume} {11}},\ \bibinfo {pages}
  {884} (\bibinfo {year} {2015})}\BibitemShut {NoStop}%
\bibitem [{\citenamefont {Geng}(2015)}]{geng15jcp}%
  \BibitemOpen
  \bibfield  {author} {\bibinfo {author} {\bibfnamefont {H.~Y.}\ \bibnamefont
  {Geng}},\ }\href@noop {} {\bibfield  {journal} {\bibinfo  {journal} {J.
  Comput. Phys.}\ }\textbf {\bibinfo {volume} {283}},\ \bibinfo {pages} {299}
  (\bibinfo {year} {2015})}\BibitemShut {NoStop}%
\bibitem [{\citenamefont {Habershon}, \citenamefont {Markland},\ and\
  \citenamefont {Manolopoulos}(2009)}]{habe+09jcp}%
  \BibitemOpen
  \bibfield  {author} {\bibinfo {author} {\bibfnamefont {S.}~\bibnamefont
  {Habershon}}, \bibinfo {author} {\bibfnamefont {T.~E.}\ \bibnamefont
  {Markland}}, \ and\ \bibinfo {author} {\bibfnamefont {D.~E.}\ \bibnamefont
  {Manolopoulos}},\ }\href@noop {} {\bibfield  {journal} {\bibinfo  {journal}
  {J. Chem. Phys.}\ }\textbf {\bibinfo {volume} {131}},\ \bibinfo {pages}
  {24501} (\bibinfo {year} {2009})}\BibitemShut {NoStop}%
\bibitem [{\citenamefont {Ceriotti}, \citenamefont {More},\ and\ \citenamefont
  {Manolopoulos}(2014)}]{ceri+14cpc}%
  \BibitemOpen
  \bibfield  {author} {\bibinfo {author} {\bibfnamefont {M.}~\bibnamefont
  {Ceriotti}}, \bibinfo {author} {\bibfnamefont {J.}~\bibnamefont {More}}, \
  and\ \bibinfo {author} {\bibfnamefont {D.~E.}\ \bibnamefont {Manolopoulos}},\
  }\href@noop {} {\bibfield  {journal} {\bibinfo  {journal} {Comput. Phys.
  Commun.}\ }\textbf {\bibinfo {volume} {185}},\ \bibinfo {pages} {1019}
  (\bibinfo {year} {2014})}\BibitemShut {NoStop}%
\bibitem [{\citenamefont {VandeVondele}\ \emph {et~al.}(2005)\citenamefont
  {VandeVondele}, \citenamefont {Krack}, \citenamefont {Mohamed}, \citenamefont
  {Parrinello}, \citenamefont {Chassaing},\ and\ \citenamefont
  {Hutter}}]{vand+05cpc}%
  \BibitemOpen
  \bibfield  {author} {\bibinfo {author} {\bibfnamefont {J.}~\bibnamefont
  {VandeVondele}}, \bibinfo {author} {\bibfnamefont {M.}~\bibnamefont {Krack}},
  \bibinfo {author} {\bibfnamefont {F.}~\bibnamefont {Mohamed}}, \bibinfo
  {author} {\bibfnamefont {M.}~\bibnamefont {Parrinello}}, \bibinfo {author}
  {\bibfnamefont {T.}~\bibnamefont {Chassaing}}, \ and\ \bibinfo {author}
  {\bibfnamefont {J.}~\bibnamefont {Hutter}},\ }\href@noop {} {\bibfield
  {journal} {\bibinfo  {journal} {Comput. Phys. Commun.}\ }\textbf {\bibinfo
  {volume} {167}},\ \bibinfo {pages} {103} (\bibinfo {year}
  {2005})}\BibitemShut {NoStop}%
\bibitem [{\citenamefont {Ma}, \citenamefont {Izaguirre},\ and\ \citenamefont
  {Skeel}(2003)}]{Ma-jsc51-verlet_impulse_instability}%
  \BibitemOpen
  \bibfield  {author} {\bibinfo {author} {\bibfnamefont {Q.}~\bibnamefont
  {Ma}}, \bibinfo {author} {\bibfnamefont {J.~A.}\ \bibnamefont {Izaguirre}}, \
  and\ \bibinfo {author} {\bibfnamefont {R.~D.}\ \bibnamefont {Skeel}},\
  }\href@noop {} {\bibfield  {journal} {\bibinfo  {journal} {SIAM Journal on
  Scientific Computing}\ }\textbf {\bibinfo {volume} {24}},\ \bibinfo {pages}
  {1951} (\bibinfo {year} {2003})}\BibitemShut {NoStop}%
\bibitem [{\citenamefont {Barth}\ and\ \citenamefont
  {Schlick}(1998)}]{barth_jcp98_mts_extra_vs_imp}%
  \BibitemOpen
  \bibfield  {author} {\bibinfo {author} {\bibfnamefont {E.}~\bibnamefont
  {Barth}}\ and\ \bibinfo {author} {\bibfnamefont {T.}~\bibnamefont
  {Schlick}},\ }\href@noop {} {\bibfield  {journal} {\bibinfo  {journal} {The
  Journal of Chemical Physics}\ }\textbf {\bibinfo {volume} {109}},\ \bibinfo
  {pages} {1633} (\bibinfo {year} {1998})}\BibitemShut {NoStop}%
\bibitem [{\citenamefont {Cao}\ and\ \citenamefont
  {Voth}(1994)}]{cao-voth94jcp}%
  \BibitemOpen
  \bibfield  {author} {\bibinfo {author} {\bibfnamefont {J.}~\bibnamefont
  {Cao}}\ and\ \bibinfo {author} {\bibfnamefont {G.~A.}\ \bibnamefont {Voth}},\
  }\href@noop {} {\bibfield  {journal} {\bibinfo  {journal} {J. Chem. Phys.}\
  }\textbf {\bibinfo {volume} {101}},\ \bibinfo {pages} {6168} (\bibinfo {year}
  {1994})}\BibitemShut {NoStop}%
\bibitem [{\citenamefont {Craig}\ and\ \citenamefont
  {Manolopoulos}(2004)}]{crai-mano04jcp}%
  \BibitemOpen
  \bibfield  {author} {\bibinfo {author} {\bibfnamefont {I.~R.}\ \bibnamefont
  {Craig}}\ and\ \bibinfo {author} {\bibfnamefont {D.~E.}\ \bibnamefont
  {Manolopoulos}},\ }\href@noop {} {\bibfield  {journal} {\bibinfo  {journal}
  {J. Chem. Phys.}\ }\textbf {\bibinfo {volume} {121}},\ \bibinfo {pages}
  {3368} (\bibinfo {year} {2004})}\BibitemShut {NoStop}%
\bibitem [{\citenamefont {Rossi}, \citenamefont {Ceriotti},\ and\ \citenamefont
  {Manolopoulos}(2014)}]{ross+14jcp}%
  \BibitemOpen
  \bibfield  {author} {\bibinfo {author} {\bibfnamefont {M.}~\bibnamefont
  {Rossi}}, \bibinfo {author} {\bibfnamefont {M.}~\bibnamefont {Ceriotti}}, \
  and\ \bibinfo {author} {\bibfnamefont {D.~E.}\ \bibnamefont {Manolopoulos}},\
  }\href@noop {} {\bibfield  {journal} {\bibinfo  {journal} {J. Chem. Phys.}\
  }\textbf {\bibinfo {volume} {140}},\ \bibinfo {pages} {234116} (\bibinfo
  {year} {2014})}\BibitemShut {NoStop}%
\bibitem [{\citenamefont {Morrone}\ \emph {et~al.}(2011)\citenamefont
  {Morrone}, \citenamefont {Markland}, \citenamefont {Ceriotti},\ and\
  \citenamefont {Berne}}]{morr+11jcp}%
  \BibitemOpen
  \bibfield  {author} {\bibinfo {author} {\bibfnamefont {J.~a.}\ \bibnamefont
  {Morrone}}, \bibinfo {author} {\bibfnamefont {T.~E.}\ \bibnamefont
  {Markland}}, \bibinfo {author} {\bibfnamefont {M.}~\bibnamefont {Ceriotti}},
  \ and\ \bibinfo {author} {\bibfnamefont {B.~J.}\ \bibnamefont {Berne}},\
  }\href@noop {} {\bibfield  {journal} {\bibinfo  {journal} {J. Chem. Phys.}\
  }\textbf {\bibinfo {volume} {134}},\ \bibinfo {pages} {14103} (\bibinfo
  {year} {2011})}\BibitemShut {NoStop}%
\bibitem [{\citenamefont {Ceriotti}\ \emph {et~al.}(2010)\citenamefont
  {Ceriotti}, \citenamefont {Parrinello}, \citenamefont {Markland},\ and\
  \citenamefont {Manolopoulos}}]{ceri+10jcp}%
  \BibitemOpen
  \bibfield  {author} {\bibinfo {author} {\bibfnamefont {M.}~\bibnamefont
  {Ceriotti}}, \bibinfo {author} {\bibfnamefont {M.}~\bibnamefont
  {Parrinello}}, \bibinfo {author} {\bibfnamefont {T.~E.}\ \bibnamefont
  {Markland}}, \ and\ \bibinfo {author} {\bibfnamefont {D.~E.}\ \bibnamefont
  {Manolopoulos}},\ }\href@noop {} {\bibfield  {journal} {\bibinfo  {journal}
  {J. Chem. Phys.}\ }\textbf {\bibinfo {volume} {133}},\ \bibinfo {pages}
  {124104} (\bibinfo {year} {2010})}\BibitemShut {NoStop}%
\bibitem [{\citenamefont {Rossi}\ \emph {et~al.}(2014)\citenamefont {Rossi},
  \citenamefont {Liu}, \citenamefont {Paesani}, \citenamefont {Bowman},\ and\
  \citenamefont {Ceriotti}}]{ross+14jcp2}%
  \BibitemOpen
  \bibfield  {author} {\bibinfo {author} {\bibfnamefont {M.}~\bibnamefont
  {Rossi}}, \bibinfo {author} {\bibfnamefont {H.}~\bibnamefont {Liu}}, \bibinfo
  {author} {\bibfnamefont {F.}~\bibnamefont {Paesani}}, \bibinfo {author}
  {\bibfnamefont {J.}~\bibnamefont {Bowman}}, \ and\ \bibinfo {author}
  {\bibfnamefont {M.}~\bibnamefont {Ceriotti}},\ }\href@noop {} {\bibfield
  {journal} {\bibinfo  {journal} {J. Chem. Phys.}\ }\textbf {\bibinfo {volume}
  {141}},\ \bibinfo {pages} {181101} (\bibinfo {year} {2014})}\BibitemShut
  {NoStop}%
\bibitem [{\citenamefont {{Del Ben}}, \citenamefont {Hutter},\ and\
  \citenamefont {VandeVondele}(2013)}]{delb+13jctc}%
  \BibitemOpen
  \bibfield  {author} {\bibinfo {author} {\bibfnamefont {M.}~\bibnamefont {{Del
  Ben}}}, \bibinfo {author} {\bibfnamefont {J.}~\bibnamefont {Hutter}}, \ and\
  \bibinfo {author} {\bibfnamefont {J.}~\bibnamefont {VandeVondele}},\
  }\href@noop {} {\bibfield  {journal} {\bibinfo  {journal} {J. Chem. Theory
  Comput.}\ }\textbf {\bibinfo {volume} {9}},\ \bibinfo {pages} {2654}
  (\bibinfo {year} {2013})}\BibitemShut {NoStop}%
\bibitem [{\citenamefont {{Del Ben}}\ \emph {et~al.}(2015)\citenamefont {{Del
  Ben}}, \citenamefont {Sch{\"{u}}tt}, \citenamefont {Wentz}, \citenamefont
  {Messmer}, \citenamefont {Hutter},\ and\ \citenamefont
  {VandeVondele}}]{delb+15cpc}%
  \BibitemOpen
  \bibfield  {author} {\bibinfo {author} {\bibfnamefont {M.}~\bibnamefont {{Del
  Ben}}}, \bibinfo {author} {\bibfnamefont {O.}~\bibnamefont {Sch{\"{u}}tt}},
  \bibinfo {author} {\bibfnamefont {T.}~\bibnamefont {Wentz}}, \bibinfo
  {author} {\bibfnamefont {P.}~\bibnamefont {Messmer}}, \bibinfo {author}
  {\bibfnamefont {J.}~\bibnamefont {Hutter}}, \ and\ \bibinfo {author}
  {\bibfnamefont {J.}~\bibnamefont {VandeVondele}},\ }\href@noop {} {\bibfield
  {journal} {\bibinfo  {journal} {Comput. Phys. Commun.}\ }\textbf {\bibinfo
  {volume} {187}},\ \bibinfo {pages} {120} (\bibinfo {year}
  {2015})}\BibitemShut {NoStop}%
\bibitem [{Note1()}]{Note1}%
  \BibitemOpen
  \bibinfo {note} {To achieve an effective load balancing, if $Q$ is the number
  of processes assigned to each of the $P$ DFT tasks, and $C_\protect \text
  {MP2}$ and $C_\protect \text {DFT}$ are the scalar costs of each force
  evaluation at the two levels of theory, each of the contracted MP2 processes
  should be assigned about $P Q C_\protect \text {MP2}/(m(C_\protect \text
  {MP2}+C_\protect \text {DFT}))$ processes, and $P Q C_\protect \text
  {DFT}/(m(C_\protect \text {MP2}+C_\protect \text {DFT}))$ processes should be
  given to each contracted DFT task.}\BibitemShut {Stop}%
\bibitem [{\citenamefont {Perdew}, \citenamefont {Burke},\ and\ \citenamefont
  {Ernzerhof}(1996)}]{perd+96prl}%
  \BibitemOpen
  \bibfield  {author} {\bibinfo {author} {\bibfnamefont {J.~P.}\ \bibnamefont
  {Perdew}}, \bibinfo {author} {\bibfnamefont {K.}~\bibnamefont {Burke}}, \
  and\ \bibinfo {author} {\bibfnamefont {M.}~\bibnamefont {Ernzerhof}},\
  }\href@noop {} {\bibfield  {journal} {\bibinfo  {journal} {Phys. Rev. Lett.}\
  }\bibinfo {series} {Phys. Rev. Lett. (USA)},\ \textbf {\bibinfo {volume}
  {77}},\ \bibinfo {pages} {3865} (\bibinfo {year} {1996})}\BibitemShut
  {NoStop}%
\bibitem [{\citenamefont {{Del Ben}}, \citenamefont {Hutter},\ and\
  \citenamefont {Vandevondele}(2012)}]{delb+12jctc}%
  \BibitemOpen
  \bibfield  {author} {\bibinfo {author} {\bibfnamefont {M.}~\bibnamefont {{Del
  Ben}}}, \bibinfo {author} {\bibfnamefont {J.}~\bibnamefont {Hutter}}, \ and\
  \bibinfo {author} {\bibfnamefont {J.}~\bibnamefont {Vandevondele}},\
  }\href@noop {} {\bibfield  {journal} {\bibinfo  {journal} {J. Chem. Theory
  Comput.}\ }\textbf {\bibinfo {volume} {8}},\ \bibinfo {pages} {4177}
  (\bibinfo {year} {2012})}\BibitemShut {NoStop}%
\bibitem [{\citenamefont {{Del Ben}}, \citenamefont {Hutter},\ and\
  \citenamefont {VandeVondele}(2015{\natexlab{b}})}]{delb+15jcp}%
  \BibitemOpen
  \bibfield  {author} {\bibinfo {author} {\bibfnamefont {M.}~\bibnamefont {{Del
  Ben}}}, \bibinfo {author} {\bibfnamefont {J.}~\bibnamefont {Hutter}}, \ and\
  \bibinfo {author} {\bibfnamefont {J.}~\bibnamefont {VandeVondele}},\
  }\href@noop {} {\bibfield  {journal} {\bibinfo  {journal} {J. Chem. Phys.}\
  }\textbf {\bibinfo {volume} {143}},\ \bibinfo {pages} {102803} (\bibinfo
  {year} {2015}{\natexlab{b}})}\BibitemShut {NoStop}%
\bibitem [{\citenamefont {Goedecker}, \citenamefont {Teter},\ and\
  \citenamefont {Hutter}(1996)}]{goed-physrevb-sep_sdsgpp}%
  \BibitemOpen
  \bibfield  {author} {\bibinfo {author} {\bibfnamefont {S.}~\bibnamefont
  {Goedecker}}, \bibinfo {author} {\bibfnamefont {M.}~\bibnamefont {Teter}}, \
  and\ \bibinfo {author} {\bibfnamefont {J.}~\bibnamefont {Hutter}},\ }\href
  {\doibase 10.1103/PhysRevB.54.1703} {\bibfield  {journal} {\bibinfo
  {journal} {Phys. Rev. B}\ }\textbf {\bibinfo {volume} {54}},\ \bibinfo
  {pages} {1703} (\bibinfo {year} {1996})}\BibitemShut {NoStop}%
\bibitem [{\citenamefont {Dahlke}, ,\ and\ \citenamefont
  {Truhlar*}(2005)}]{dahlk-jpcb-improved_functionals_h2o}%
  \BibitemOpen
  \bibfield  {author} {\bibinfo {author} {\bibfnamefont {E.~E.}\ \bibnamefont
  {Dahlke}}, , \ and\ \bibinfo {author} {\bibfnamefont {D.~G.}\ \bibnamefont
  {Truhlar*}},\ }\href@noop {} {\bibfield  {journal} {\bibinfo  {journal} {The
  Journal of Physical Chemistry B}\ }\textbf {\bibinfo {volume} {109}},\
  \bibinfo {pages} {15677} (\bibinfo {year} {2005})}\BibitemShut {NoStop}%
\bibitem [{\citenamefont {Grimme}\ \emph {et~al.}(2010)\citenamefont {Grimme},
  \citenamefont {Antony}, \citenamefont {Ehrlich},\ and\ \citenamefont
  {Krieg}}]{grimme-jcp10-dftd_disp_functional}%
  \BibitemOpen
  \bibfield  {author} {\bibinfo {author} {\bibfnamefont {S.}~\bibnamefont
  {Grimme}}, \bibinfo {author} {\bibfnamefont {J.}~\bibnamefont {Antony}},
  \bibinfo {author} {\bibfnamefont {S.}~\bibnamefont {Ehrlich}}, \ and\
  \bibinfo {author} {\bibfnamefont {H.}~\bibnamefont {Krieg}},\ }\href@noop {}
  {\bibfield  {journal} {\bibinfo  {journal} {The Journal of Chemical Physics}\
  }\textbf {\bibinfo {volume} {132}},\ \bibinfo {pages} {154104} (\bibinfo
  {year} {2010})}\BibitemShut {NoStop}%
\bibitem [{Note2()}]{Note2}%
  \BibitemOpen
  \bibinfo {note} {The development version we used here is provided in the SI.
  More recent versions can be obtained from the Authors.}\BibitemShut {Stop}%
\bibitem [{\citenamefont {Marsakel}\ and\ \citenamefont {Markland}()}]{TE-PC}%
  \BibitemOpen
  \bibfield  {author} {\bibinfo {author} {\bibfnamefont {O.}~\bibnamefont
  {Marsakel}}\ and\ \bibinfo {author} {\bibfnamefont {T.}~\bibnamefont
  {Markland}},\ }\href@noop {} {}\bibinfo {howpublished} {private
  communication}\BibitemShut {NoStop}%
\end{thebibliography}
\end{document}